\documentclass[twocolumn,showpacs,pra,amsmath,amssymb]{revtex4}

\usepackage{graphicx}
\usepackage{dcolumn}
\usepackage{bm}

\usepackage[matrix,frame,arrow]{xy}

\begin{document}

\title{Efficient quantum circuit implementation of quantum walks}
\author{B L Douglas}
\author{J B Wang}

\affiliation{School of Physics, The University of Western Australia, 6009, Perth, Australia.}

\begin{abstract}

Quantum walks, being the quantum analogue of classical random walks, are expected to provide a fruitful source of quantum algorithms. A few such algorithms have already been developed, including the `glued trees' algorithm, which provides an exponential speedup over classical methods, relative to a particular quantum oracle. Here, we discuss the possibility of a quantum walk algorithm yielding such an exponential speedup over possible classical algorithms, without the use of an oracle. We provide examples of some highly symmetric graphs on which efficient quantum circuits implementing quantum walks can be constructed, and discuss potential applications to quantum search for marked vertices along these graphs.

\end{abstract}

\pacs{03.67.Lx, 03.67.-a, 89.20.Ff}

\maketitle

\section{Introduction}

The considerable, ongoing interest in quantum algorithms has been sparked by the possibility of practical solutions to problems that cannot be efficiently solved by classical computers. In other words, the opportunity to achieve exponential speedups over classical techniques by harnessing entanglement between densely encoded states in a quantum computer. Quantum walks have been the focus of several recent studies (see for example, \cite{kempe, general1, general2, general3, general4}), with particular interest in possible algorithmic applications of the walks \cite{childs, app1, hypercube, app3, app4}. A few such algorithms have already been developed, perhaps the most notable being the `glued trees' algorithm developed by Childs \textit{et al}. \cite{childs}, in which quantum walks are shown to traverse a family of graphs exponentially faster than any possible classical algorithm, given a certain quantum oracle.

In this paper we discuss the possibility of a quantum walk algorithm providing such an exponential speedup over possible classical algorithms without the use of an oracle. Firstly, we present a formal construction of quantum walks, and show that they can be implemented classically in a time that scales polynomially with the size of the state space. We then consider an efficient quantum implementation of quantum walks to be one in which the resources required scales logarithmically with the size of the state space, and present examples of graphs for which such an implementation is possible.

\section{Quantum random walks}

Quantum walks can be thought of as the quantum analogue of simple classical random walks. They are a unitary process, and can be naturally implemented by quantum systems. The discrete-time walk consists of a unitary operator $U = S C$, where $S$ and $C$ are termed the shifting and coin operators respectively, acting on the state space.

Consider a discrete-time quantum walk along a general undirected graph $G(V,E)$, with vertex set $V=\left\{v_1,v_2,v_3,\ldots\right\}$, and edge set $E=\left\{(v_i,v_j),(v_k,v_l),\ldots\right\}$, being unordered pairs connecting the vertices. The quantum walk acts on an extended position space, in which each node $v_i$ with valency $d_i$ is split into $d_i$ subnodes. This space then consists of all states $(v_i, a_i)$, where $v_i \in V$ and $1 \le a_i \le d_i$. The shifting operator acts on this extended position space, with its action defined by:
$$S(v_i,a_i) = (v_j,a_j),$$
for some $v_j \in V$ such that $(v_i, v_j) \in E$. The coin operator comprises a group of unitary operators, or a set of coins, each of which independently mix the probability amplitudes associated with the group of sub-nodes of a given node. For example, given a vertex $v_i$ with valency $d_i$, the coin can be represented by a unitary $(d_i \times d_i)$ matrix.

This definition is necessarily vague, allowing significant freedom in the construction of shifting and coin operators, depending on the desired properties. If, for example, a specific labeling of the vertices of the graph was not known, the shifting and coin operators may be required to act symmetrically with respect to any arbitrary labeling. This means that the coin matrix must be symmetric, and the shifting can take place only along edges, with $S^2$ equaling the identity operator.

Consider an undirected graph, having order $n$ and $k$ edges, with no self loops or multiple edges between pairs of vertices. Then the above definition yields a state space with $2k$ states. The shifting operator $S$ can then be represented by a $(2k \times 2k)$ permutation matrix, and if we group the states derived from a common vertex, the coin operator $C$ can be represented by a $(2k \times 2k)$ block diagonal matrix. Since $k$ has an upper bound of $n(n-1)/2$, it follows that a step of the walk, $U=S C$, can be simulated efficiently on a classical computer, in a time that scales with $O(n^6)$. In fact, the shifting operator, being a permutation of the $2k$ states, can be implemented more efficiently with an upper bound scaling of $O(n^4)$ \cite{GIpaper}, as can the coin operator, containing $n$ blocks of size at most $n$. Hence, quantum walks on graphs can be classically simulated in polynomially time, scaling with graph size. So for even the possibility of exponential speedups, quantum implementations must scale logarithmically with graph size.

Many of the currently proposed `natural' physical implementations of quantum walks \cite{impl1,impl2,impl3,impl4} cannot achieve this, as the walks evolve on nodes that are implemented by physical states, on which operations are directly performed. Hence the resource requirements grow polynomially with the state space. In order to achieve an exponential gain, the nodes need to foremost be encoded by a string of entangled states, such as qubits in a quantum computer, making use of memory that grows exponentially with the number of qubits. In addition, the number of elementary gates required to perform the walk needs also grow logarithmically with the size of the state space. 

So far, this has only been found to be possible for structures with a high degree of symmetry - where symmetry in this case refers to the ability to characterize the structure by a small number of parameters, increasing at most logarithmically with the number of nodes. Note that this may not necessarily imply that the structure has geometric or combinatorial symmetry in the typical sense of the terms. For instance, sparse graphs with efficiently computable neighbors fall into this category, and as a consequence of \cite{sparse2, sparse3}  have been shown to allow efficient implementations of quantum walks. Here sparse graphs of order $n$ are defined as in \cite{sparse2} to have degree bounded by O(polylog($n$)), with the further condition that the neighbors of each vertex are efficiently computable. Possessing efficiently computable neighbors implies the existence of an O(log($n$)) sized function characterizing the graph, such that the information contained in the O($n$) edges can be compressed to size O(log($n$)). This compression seems to require the presence of some kind of structure to the system, for example, the graph cannot contain more than O(log($n$)) completely random edges.  An interesting open question is whether sparse graphs can have no automorphisms apart from the identity. 

\section{Efficient quantum circuit implementation}

In this section, we give examples of a few such graphs for which relatively simple quantum circuits can be designed to efficiently implement quantum walks along them.
Firstly, we will look at a simple cycle. To implement a quantum walk along it, we first note that each node has two adjacent edges, and hence two subnodes. Proceeding systematically around the cycle, we assign each node a bit-string value in lexicographic order, such that adjacent nodes are given adjacent bit-strings. For a cycle of order $2^n$, $n$ qubits are required to encode the nodes, and an additional qubit to encode the subnodes. The coin operation can be implemented by a single Hadamard gate acting on the subnode qubit, and the shifting operation by a cyclic permutation of the node states, in which each state (or bit-string) is mapped to an adjacent state (either higher or lower depending on the value of the subnode qubit).

\begin{figure}[htb]
\centerline{
\includegraphics[width=6cm]{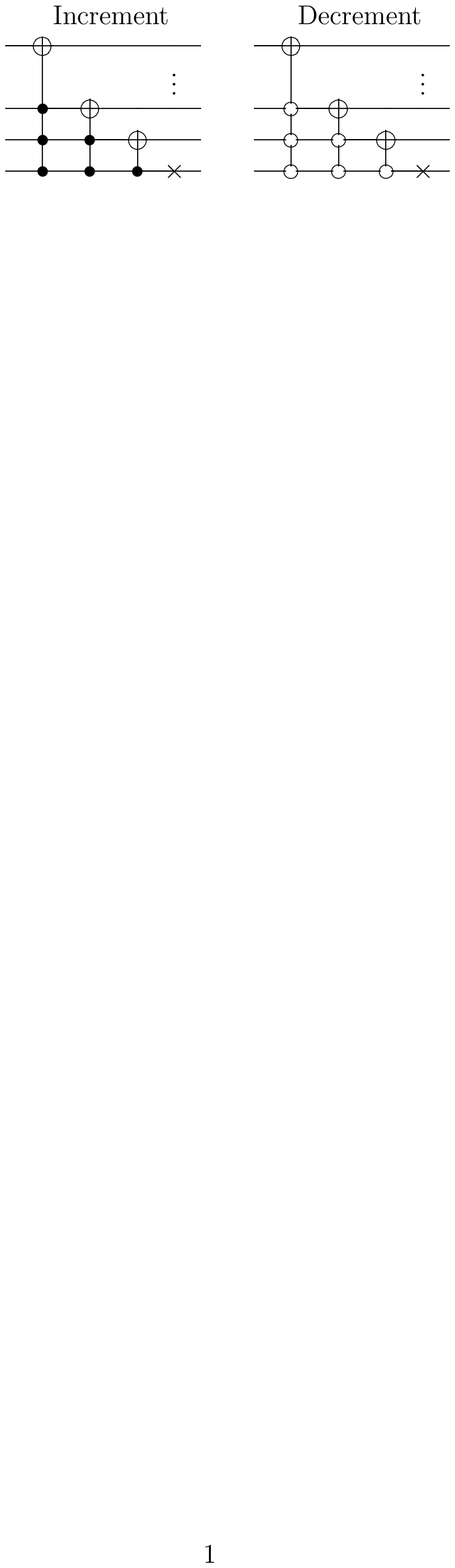} 
}
\caption{\label{incdec} Increment and decrement gates on $n$ qubits, producing cyclic permutations in the $2^n$ bit-string states.}
\end{figure}

\begin{figure}[htb]
\centerline{
\includegraphics[width=8cm]{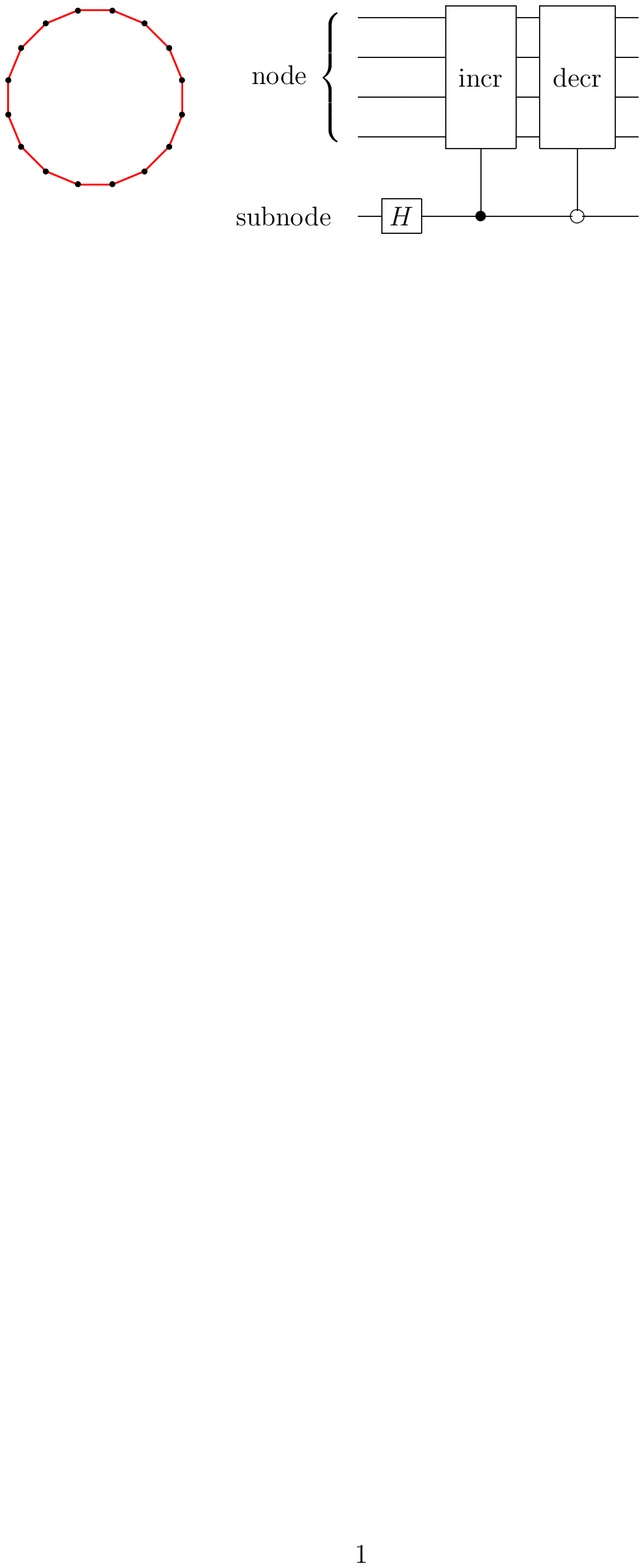} 
}
\caption{\label{circuit_cycle} Quantum circuit implementing a quantum walk along a 16-length cycle.}
\end{figure}

\begin{figure}[htb]
\centerline{
\includegraphics[width=9cm]{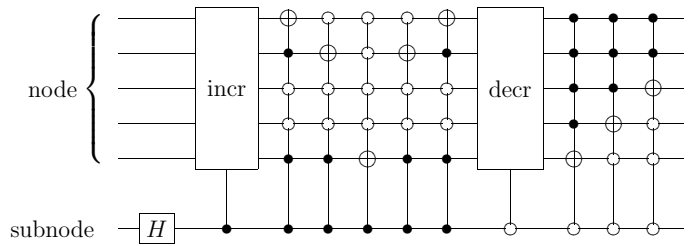} 
}
\caption{\label{cycle_25} Quantum circuit implementing a quantum walk along a 25-length cycle.}
\end{figure}

This permutation can be achieved via `increment' and `decrement' gates, shown in Figure \ref{incdec}, made up of generalized CNOT gates. These gates produce cyclic permutations (in either direction) of the node states. The resulting shifting operator is $S = ( \textrm{Incr.} \otimes \left|1\right\rangle + \textrm{Decr.} \otimes \left|0\right\rangle )$. Here the tensor space description separates the node and subnode states. So to implement a walk along a cycle of size $2^n$ we require $n+1$ qubits.  $O(n)$ additional ancillary qubits may also be required for the generalized CNOT gates involved in the cyclic permutations, depending on the specific implementation used. The number of elementary gates required is limited to $O(n)$, hence both memory and resource requirements scale logarithmically with graph size. An example of the circuit for a cycle of size 16 is given in Figure \ref{circuit_cycle}. Note that although this specific implementation requires a cycle of order $2^n$, only trivial alterations are required to efficiently implement cycles of any size. For instance, an equivalent circuit for a cycle of size 25 is given in Figure \ref{cycle_25}.

\begin{figure}[htb]
\centerline{
\includegraphics[width=9cm]{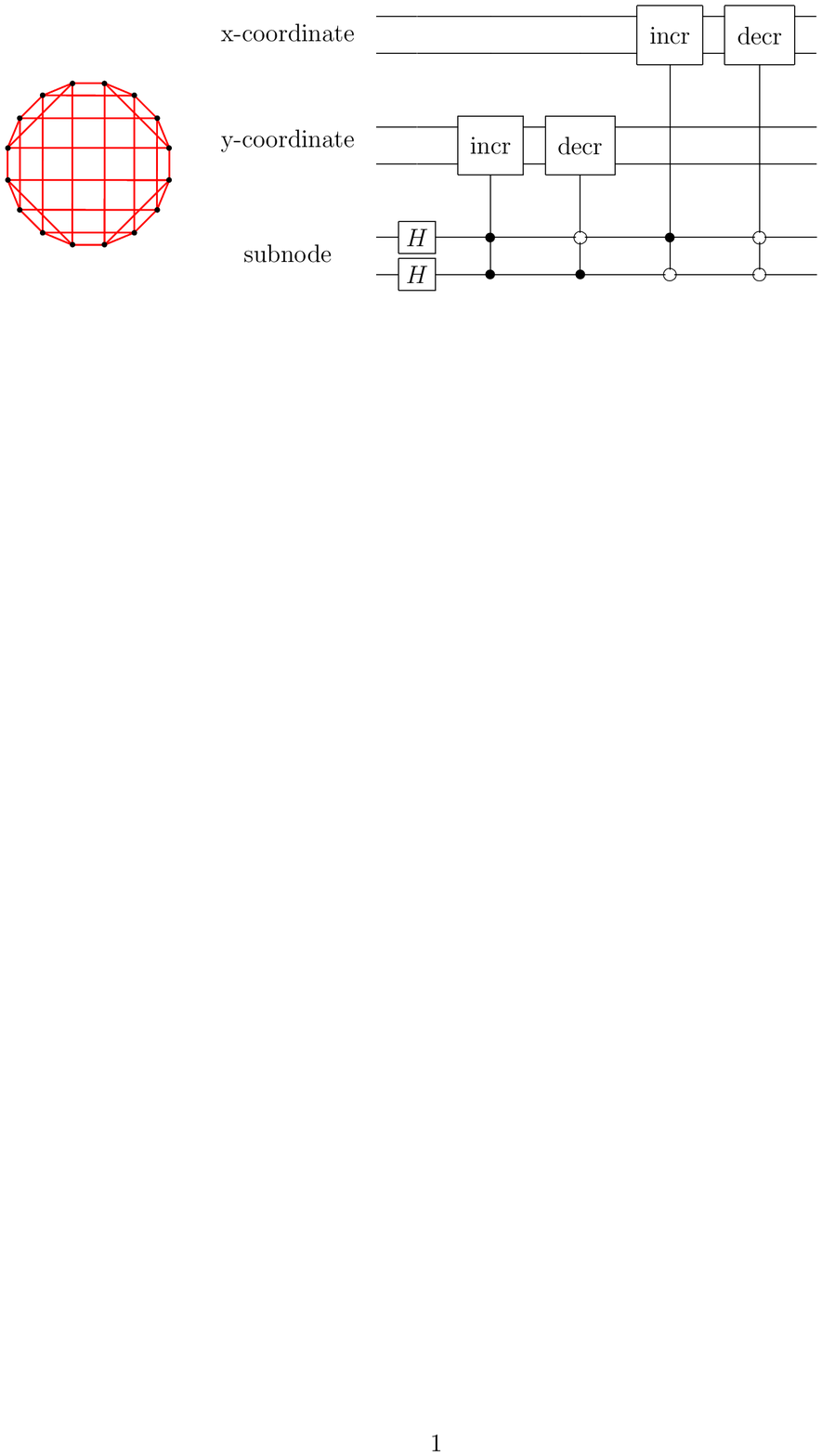} 
}
\caption{\label{circuit_grid} Quantum circuit implementing a quantum walk along a 2D hypercycle.}
\end{figure}

\begin{figure}[htb]
\includegraphics[width=4.5cm]{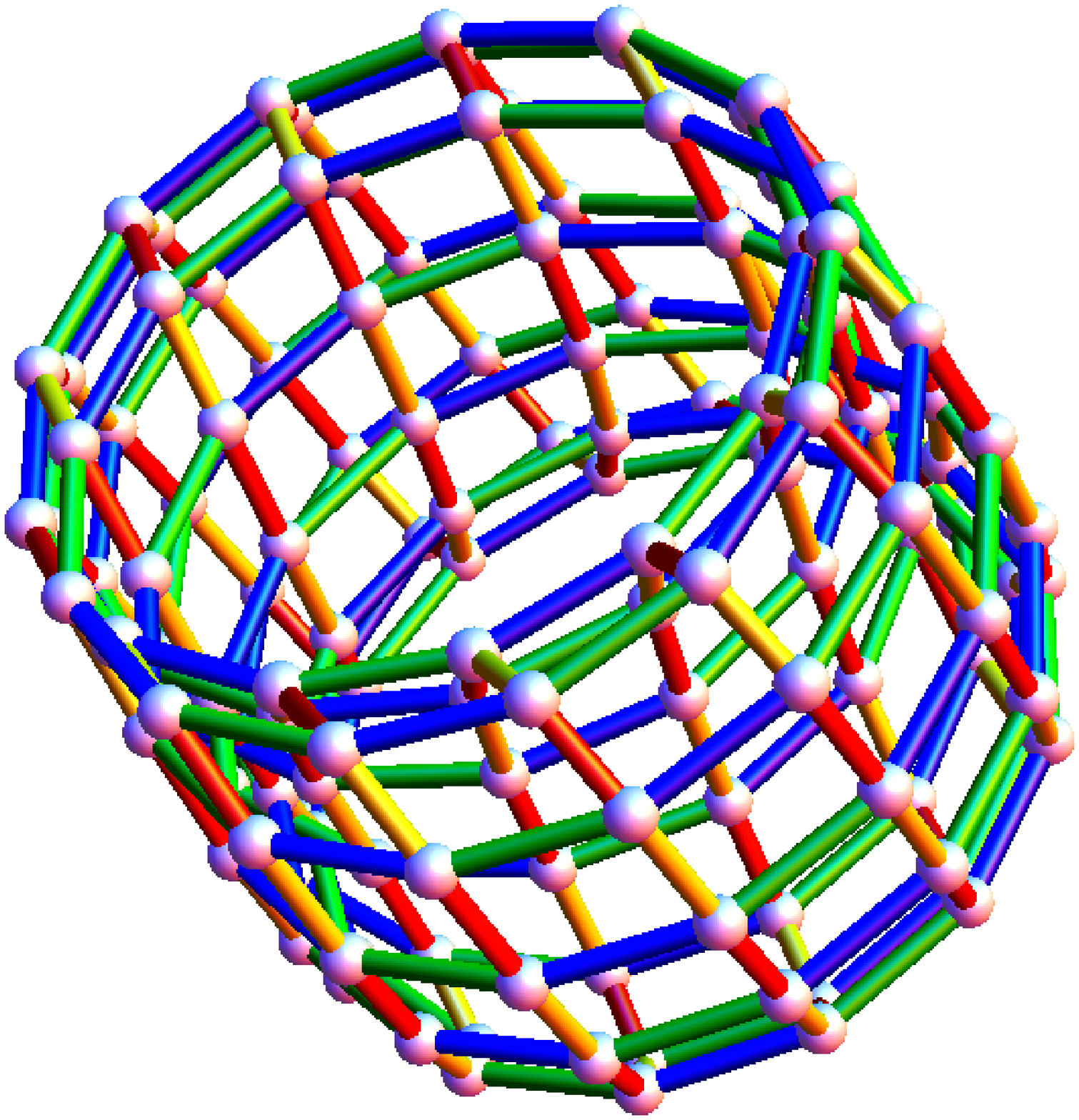}
\includegraphics[width=3.5cm]{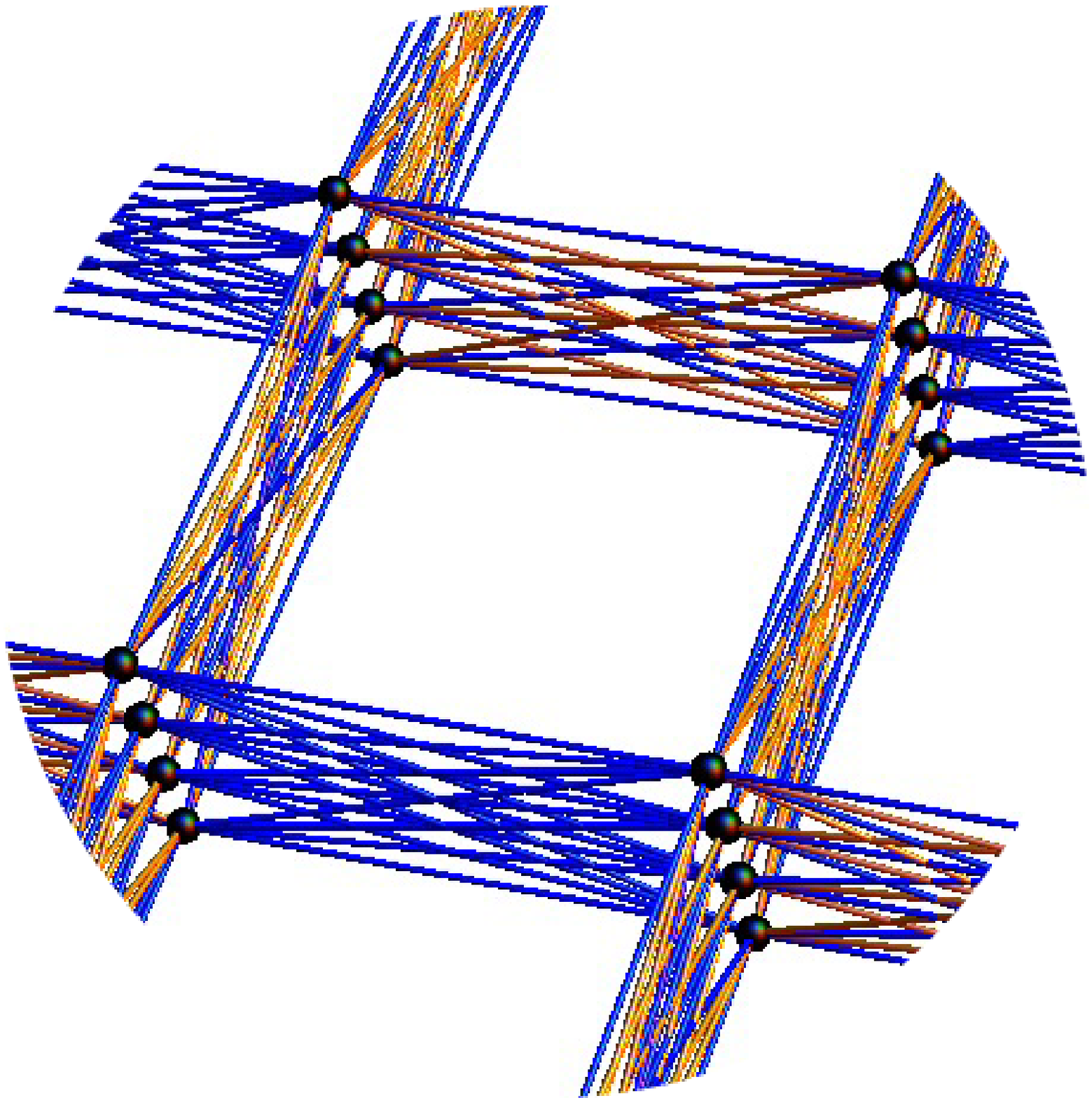}
\caption{\label{toroid} `Twisted' toroidal lattice graph. Each node in the representation on the left contains four sub-nodes of the graph, as indicated on the right.}
\end{figure}

\begin{figure}[htb]
\centerline{
\includegraphics[width=9.5cm]{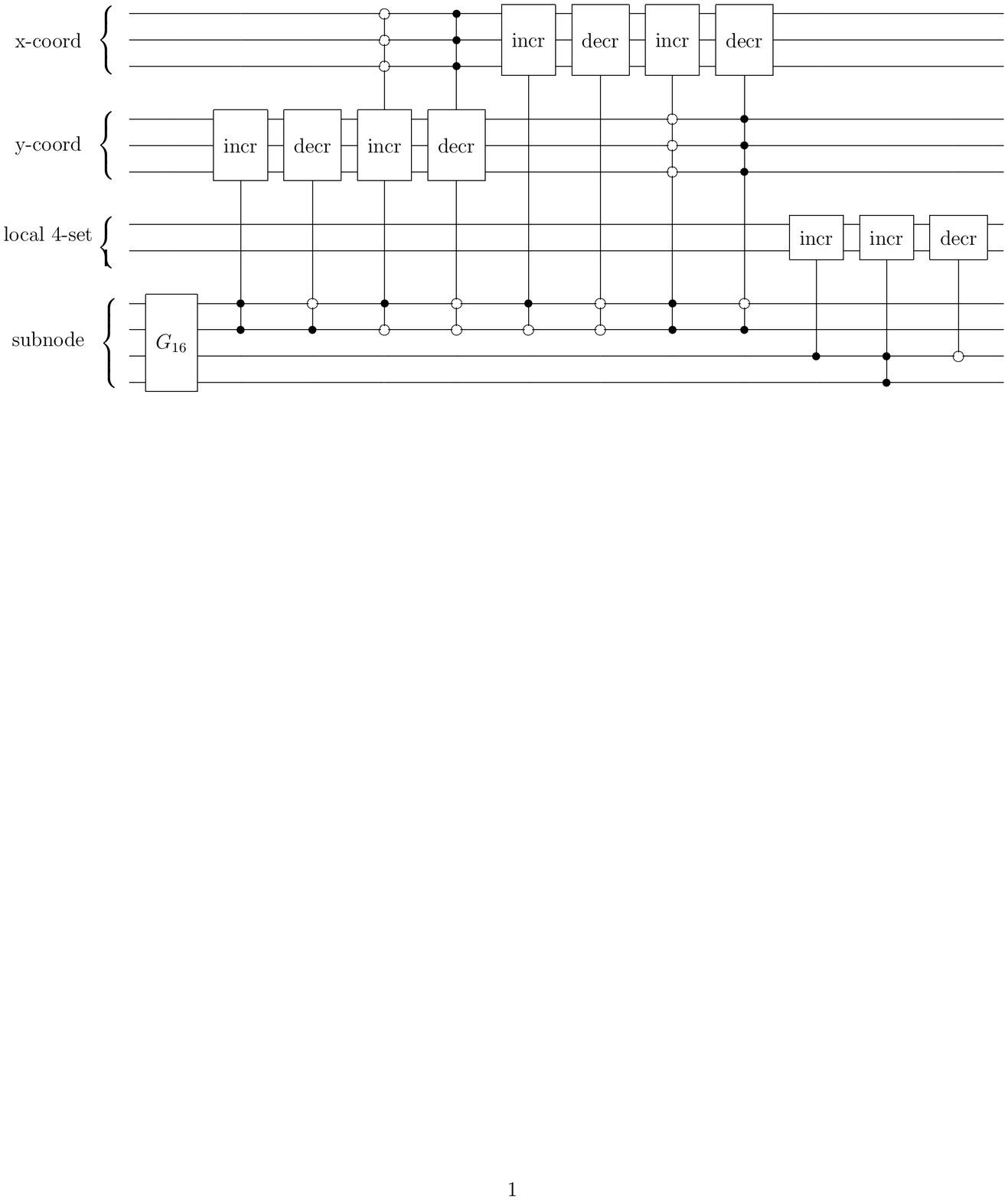} 
}
\caption{\label{circuit_toroid} Quantum circuit implementing a quantum walk along the twisted toroidal of Figure \ref{toroid}, of dimension $8 \times 8 \times 4$. }
\end{figure}

A similar method can be used to efficiently implement a walk along a $2^n$ dimensional grid or hypercube, by partitioning the labels of the nodes into $n$ distinct sets, corresponding to each coordinate. An example for the 2D ($4 \times 4$) hypercycle is given in Figure \ref{circuit_grid}. As an extension, a quantum circuit implementing a walk along a twisted toroidal supergraph as shown in Figure \ref{toroid} is given in Figure \ref{circuit_toroid}. This structure was employed by Menicucci \textit{et al} \cite{Menicucci08} to set up QC-universal toroidal lattice cluster states.

\begin{figure}[htb]
\centerline{
\includegraphics[width=9cm]{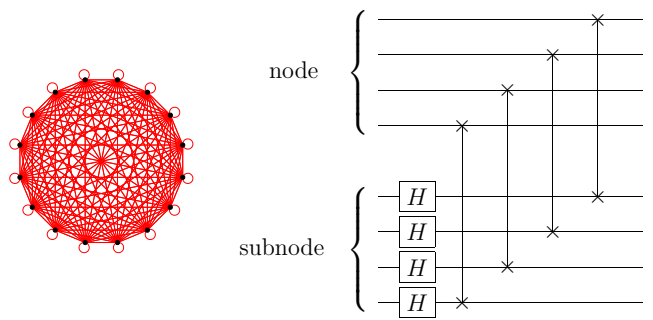} 
}
\caption{\label{circuit_complete} Quantum circuit implementing a quantum walk along a complete 16-graph.}
\end{figure}

\begin{figure}[htb]
\centerline{
\includegraphics[width=8.5cm]{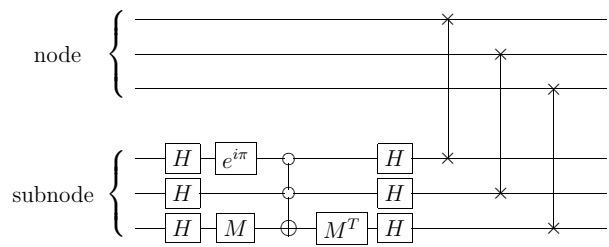} 
}
\caption{\label{complete_2_grover} Quantum circuit implementing a quantum walk along a complete 8-graph, using a Grover coin.}
\end{figure}

Other highly symmetric structures, such as the complete $2^n+1$ graph, a complete $2^n$ graph with self loops and a binary tree also allow efficient implementations of quantum walks with a qubit based quantum circuit. Walking along the complete $2^n$ graph, using the Hadamard coin operator, can be naturally implemented using only single qubit gates and CNOT gates ($n$ Hadamard gates and $3n$ CNOT gates). The circuit for a complete $2^n$ graph ($n=3$), in which each node has a self loop, is shown in Figure \ref{circuit_complete}, and is fairly intuitive.  
Alternatively, if the Grover coin operator is used, $n+3$ extra single qubit gates, one extra $\textrm{C}^{n-1}\textrm{NOT}$ gate (which is a generalized CNOT gate with $n-1$ control bits and one target bit), and $n$ Hadamard gates are required, as shown in Figure \ref{complete_2_grover}. Here 
\begin{displaymath}
 M = \frac{1}{\sqrt{2}}
\left( \begin{array}{cc}
        1 & 1 \\
	-1 & 1
       \end{array} \right)
\end{displaymath}
is a permutation of the Hadamard operator. Note that even using the Grover coin $\textrm{G}_n$, the coin operator is still mostly separable, requiring only single qubit operators apart from the one $\textrm{C}^{n-1}\textrm{NOT}$ gate.

\begin{figure}[htb]
\centerline{
\includegraphics[width=7cm]{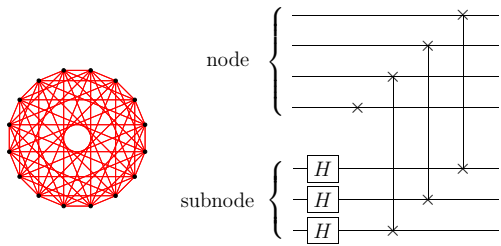} 
}
\caption{\label{complete_variant} Quantum circuit implementing a quantum walk along a complete 16-graph with every second edge removed.}
\end{figure}

Walks along highly symmetric variants of the complete graph (as opposed to sparse graphs, such as those considered previously) can also be efficiently implemented. For instance if we consider the complete graph on $2^n$ vertices, together with an arbitrary labeling of the nodes from 1 to $2^n$. Removing edges between nodes whose labels differ by a multiple of 2 leads to a regular graph of degree $2^{n-1}$, shown in Figure \ref{complete_variant} for $n = 3$. This is then a complete bipartite graph, and a walk along such a graph can be efficiently implemented by the circuit of Figure \ref{complete_variant}, an even simpler circuit than for the complete $2^n$ graph.

\begin{figure}[htb]
\includegraphics[width=7.0cm]{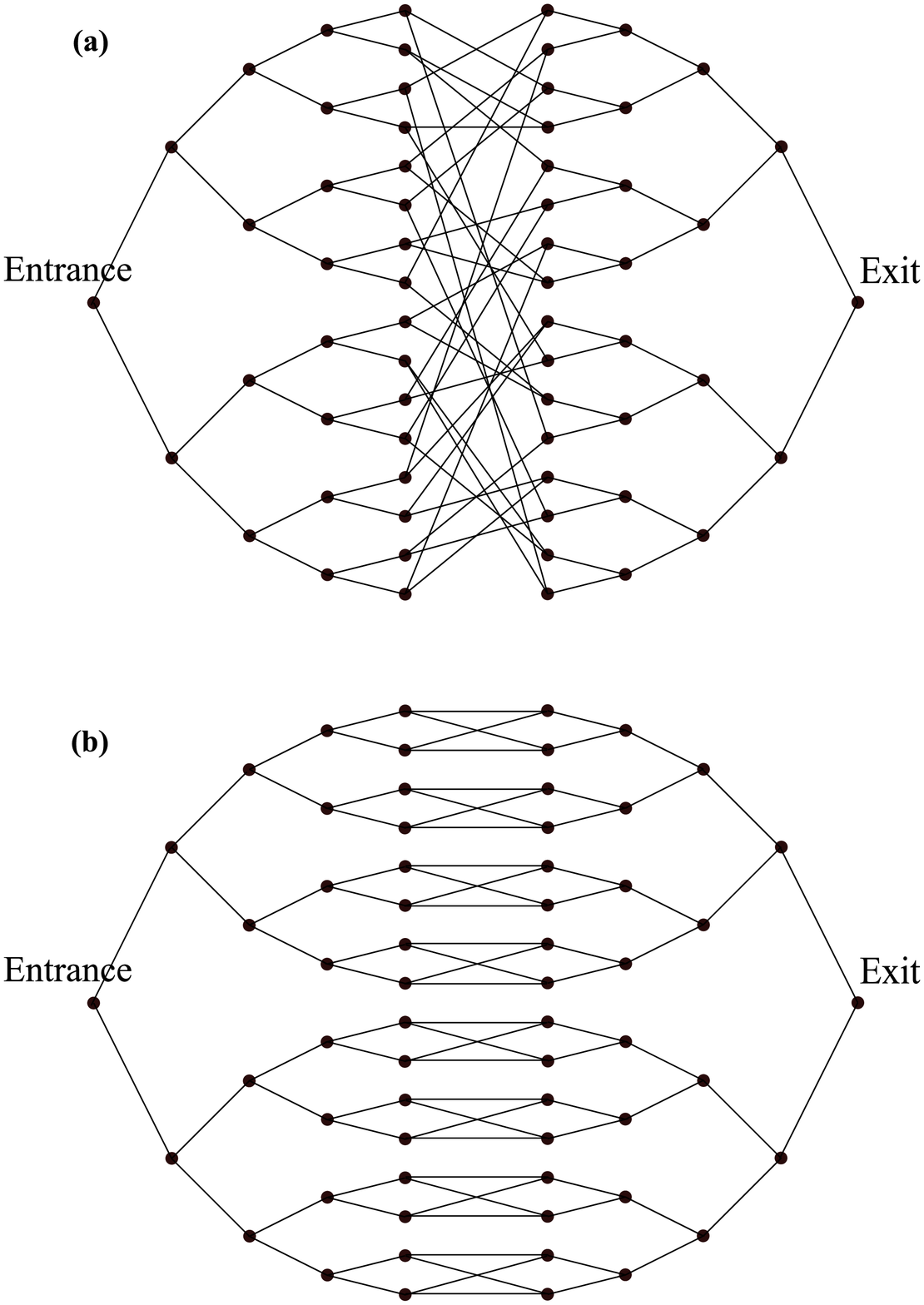}
\caption{\label{glued_tree} Binary glued trees with random (a) and regular (b) interconnections between the central levels.}
\end{figure}

\begin{figure}[htb]
\centerline{
\includegraphics[width=9cm]{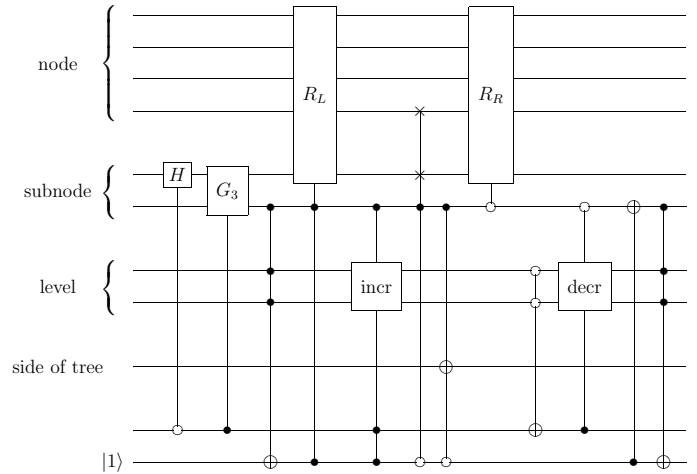} 
}
\caption{\label{circuit_tree} Quantum circuit implementing a quantum walk along a glued tree with a regular labeling of the nodes.}
\end{figure}

Given the results of Childs \textit{et al}. \cite{childs} and Cleve \textit{et al}. \cite{cleve09}, in which quantum walks are shown to traverse a family of `glued trees' exponentially faster than any possible classical algorithm, relative to a quantum oracle, we decided to look into quantum walks along glued trees in the non-oracular setting. Note that the algorithm presented in \cite{childs} employs continuous-time quantum walks, while in \cite{cleve09} it was shown to also be implementable by discrete-time quantum walks. Both require the use of a quantum oracle. In the non-oracular case, efficient implementation of a quantum walk along the glued trees is not possible given random interconnections between the central levels (as in Figure \ref{glued_tree}(a)), since this would be equivalent to performing a random permutation of $2^n$ states in time $O(\textrm{poly}(n))$. Instead we are restricted to considering regular interconnections, such as those of Figure \ref{glued_tree}(b). Here `regular' interconnections are those that can be completely characterized by $O(\textrm{poly}(n))$ parameters. The algorithm of \cite{childs} requires a symmetric coin operator - hence we use the Grover coin, defined on $d$ dimensions by $(G_d)_{i,j} = \frac{2}{d} - \delta_{i,j}$, the only purely real symmetric coin \cite{GIpaper}. We also restrict the shifting operator to $S^2 = I$, where $I$ is the identity operator. In this case an efficient quantum circuit can be constructed, for example that of Figure \ref{circuit_tree}, for tree depth 4 (with 62 nodes). Here the $G_3$ gate represents a three dimensional Grover coin operator acting on two qubits (mixing three of the four states, while the fourth is not accessed). For a tree depth of $l$, the circuit requires $l + \textrm{log}_2 l + 5$ qubits, together with $O(l)$ elementary gates.

\begin{figure}[htb]
\centerline{
\includegraphics[width=8cm]{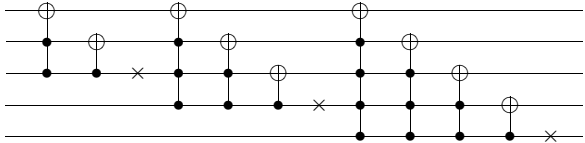} 
}
\caption{\label{rotation} A rotation of 7 states, split into the composite powers of 2, being three rotations of size 4, 2 and 1 states respectively.}
\end{figure}

\begin{figure}[htb]
\centerline{
\includegraphics[width=3cm]{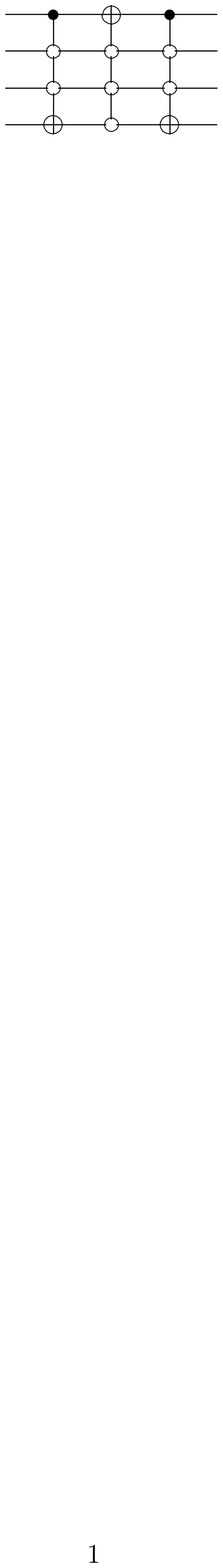}
}
\caption{\label{transpose} A transposition of the $\left|0000\right\rangle$ and $\left|1001\right\rangle$ states.}
\end{figure}

Related to the problem of which structures quantum walks can be efficiently implemented on is the question of which permutations of a set of states can be efficiently implemented. Given a set of $n$ qubits encoding $2^n$ quantum states, we wish to know which permutations of these states can be implemented using $O(\rm{poly}(n))$ elementary gate operations.  Cyclic rotations of the states (relative to the lexicographic order of their bit-strings) can be implemented efficiently, as shown above. In fact any rotation of the states can be performed efficiently, by first decomposing it into a series of rotations of size $2^m$, for some integer $m$. For instance, an incremental rotation of 7 states applied to the 32 states represented by 5 qubits is explicitly shown in Figure \ref{rotation}. Generalized control-not operations can also be used to transpose pairs of states differing in label by a single qubit. Similarly, any two states differing by $m$ qubits can be efficiently transposed using $2m-1$ generalized CNOT operations. For example, given 16 states encoded by 4 qubits, the lexicographically 1st and 10th states (represented by $\left|0000\right\rangle$ and $\left|1001\right\rangle$ respectively) can be transposed via 3 controlled swap operations, as shown in Figure \ref{transpose}. Using this method any transposition of states on $n$ qubits can be performed using a maximum of $2n -1$ generalized CNOT gates, or $2n^2-3n$ $\textrm{C}^2\textrm{NOT}$ gates. This may not be the optimal way to implement a particular transposition, however it does scale logarithmically with the number of states.

Using similar methods, other permutations with essentially binary characters can also be efficiently implemented, such as swapping every second state, or performing some given internal permutation to each consecutive group of 8 states (or $2^m$ states, for some fixed integer $m$). Note that permutations which may not seem to have a binary character can be transformed to efficiently implementable permutations. For instance, if we wished to split the set of states into groups of 6, and swap every 4th element, we could achieve this by expanding the state space - embedding each group of 6 into a group of 8, with the last two states remaining unused, `empty states'.

\begin{figure}[h]
\centerline{
\includegraphics[width=8cm]{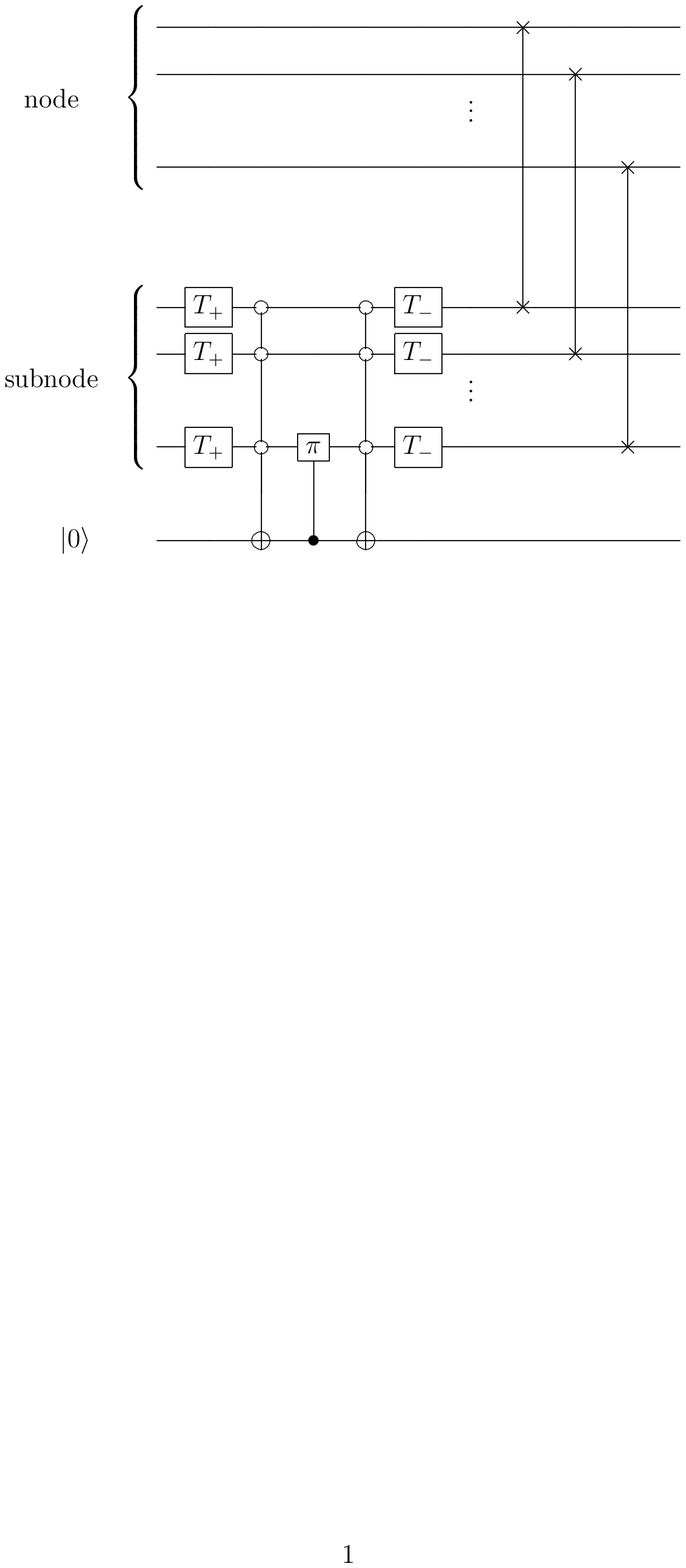} 
}
\caption{\label{complete_3} Qutrit-based quantum circuit implementing a quantum walk along a complete $3^n$-graph.}
\end{figure}

For simplicity, given an implementation on qubits, the preceding examples have all been essentially binary in nature. Efficient implementations using qubits are equally possible on other, non-binary structures, such as ternary trees or complete $3^n$ graphs. For example, implementing the complete $3^n$ graph (with self-loops) using a qubit circuit requires many more 2-qubits gates, given the need to approximate a 9D Hadamard or Grover coin operator over 16 states, without mixing into the other 7 states. As would be expected, a more natural implementation is possible if qutrits are used instead. In this case, the coin operator is again nearly separable if using the Grover coin operator, and completely separable if using a qutrit equivalent of the Hadamard operator. Here we take a qutrit equivalent of the Hadamard operator to be an operator $T_n$ acting on $n$ qutrits, satisfying:
$$((T_1)_\pm)_{a,b} = \frac{1}{\sqrt{3}} e^{\pm i \frac{2\pi}{3} a \; b} , \; \textrm{ where } \; a,b \in \{0,1,2\}, \textrm{ and}$$
$$(T_n)_\pm = (T_1)_\pm \otimes (T_{n-1})_\pm  .$$
Qutrit circuits implementing a quantum walk along the complete-$3^n$ graph using the $T$ coin operator or the Grover coin operator can then be constructed as in Figure \ref{complete_3}.
Nevertheless, the use of a more natural base still provides a polynomial efficiency gain.

\section{Conclusions}

We have presented here a set of highly symmetric graphs all amenable to exact, efficient quantum circuits implementing quantum walks along them. The examples considered here are quite simple, and more complex variations can still be efficiently implemented such as composites of highly symmetric graphs, symmetric graphs with a small number of `imperfections', as well as graphs possessing a certain bounded level of complexity.

Quantum walks have been used to search for marked vertices along highly symmetric graphs, including the hypercube, complete graphs and complete multipartite graphs \cite{reitzner, hypercube}. These studies have dealt with the computational complexity of such searches relative to an oracle - looking at the number of steps of a quantum walk required to find a marked vertex, with individual steps of the walk itself largely left to the oracle. In such cases searching using quantum walks has yielded a quadratic speedup over classical search algorithms.

In a practical implementation of such a search algorithm, the computations performed by the oracle (that is, performing a step of the walk in which the coin operator differs for marked and unmarked nodes) would of course affect the running time. The work presented here can be used to efficiently implement such an oracle - using O(log($n$)) elementary gates for a graph of order $n$ - given a highly symmetric graph such as those considered in \cite{hypercube, reitzner} and in this paper.

\begin{acknowledgments}
We thank N. Menicucci and S. Flamia for use of their Mathematica code, which was adapted to construct the toroidal graph figure used here.

\end{acknowledgments}


\end{document}